\documentclass[letterpaper,12pt]{article}   

\usepackage{graphics, epsfig}

\begin{document}

\title{\bf The nonmodular topological phase and phase singularities}

\author{\bf Rajendra~Bhandari}
\maketitle
\begin{center}
\begin{tabular}{ll}
            &19/13, 1st Main Road, \\
            & Jayamahal Extension,\\
            & Bangalore 560 046, India. \\
            &  email: rajbhand@yahoo.com\\
            &  phone: 91-80-23333658\\
\end{tabular}
\end{center}

\vspace{10mm}


\vskip 2.0cm

\newpage

\begin{abstract}

Generalizing an earlier definition of the noncyclic geometric phase (R.Bhandari, Phys.Lett.A, 157, 221 (1991)),
a nonmodular topological phase is defined with reference to a generic time-dependent 
two-slit interference experiment involving particles with N internal states in which the internal
state of both the beams undergoes unitary evolution. A simple proof of the shorter geodesic rule for 
closure of the open path is presented and several 
useful new insights into the behaviour of the dynamical and geometrical components of the 
phase shift presented. An effective hamiltonian interpretation of the observable phase shifts 
is also presented.

\end{abstract}

\maketitle
\section{Introduction}

Study of interference of particles with internal degrees of freedom e.g. a photon with its 
two polarization states and a neutron with its two spin-1/2 states has been of interest since long. 
The famous $4\pi$ spinor symmetry experiments with unpolarized neutrons of Rauch et al. \cite{rauch}
and Werner et al. \cite{werner} are some notable early examples. 
In optics, the Young's double slit experiment of Pescetti \cite{pescetti} with polarized light 
and its interpretation in terms of the measurement problem in 
quantum mechanics is another interesting early example.

The overall phase of the state of an evolving quantum system with N internal states is nontrivial when 
a possibility of interference of the state with a copy of the unevolved state (produced 
with the use of a beamsplitter), or with a copy which has evolved differently, exists. 
The simplest example would be a beam of neutrons, with its two spin states, passing through 
a box containing an arbitrarily oriented magnetic field and interfering with a part of the 
beam which has either travelled through free space or through a  box contining a different
magnetic field. As is now widely recognized, the first definition of a phase difference between beams in two 
different internal states in such a system  was given by Pancharatnam  
in the context of 
the two-state system of polarization of light.   Pancharatnam made two 
important contributions to the physics of the phase \cite{panch1}. Firstly he proposed that
the phase difference between two waves in different polarization states is the phase 
that must be introduced so that their superposition yields maximum intensity. Secondly Pancharatnam  
showed that if two states A and B are in phase 
by the above definition, A and C are in phase, then B and C have a phase difference 
equal to half the solid angle subtended at the centre by the geodesic triangle ABC on 
the sphere representing the states of the two-state system. This constituted the first formulation 
of the geometric phase. Following the discovery of the geometric phase in the context of 
adiabatic cyclic quantum evolutions by Berry in 1983 \cite{berry}, this quantity 
is sometimes called ``Berry phase". The next  step forward in the problem was 
the work of Aharonov and Anandan \cite{aa} who showed that in an evolution where the 
state evolves cyclically, if a quantity called the dynamical phase which is equal to the 
time integral of $<H(t)>$; $H(t)$ being the hamiltonian of the system, is subtracted 
from the total phase, the remainder is independent of 
the hamiltonian of the system and is the same, modulo $2\pi$, for all evolutions involving the 
same cyclic path in the state space. For a two-state system they showed that this quantity 
is equal to half the solid angle subtended by the closed path at the centre of the state sphere i.e. it is
a geometric phase .

Cyclic evolutions however form a very small sub-set of the complete set of evolutions in nature. 
Even if the evolution is cyclic one may wish to consider only a part of the cycle. To define 
the phase of a state under a noncyclic evolution or for a partial cycle one needs to 
take recourse to Pancharatnam's criterion mentioned above. The first attempt to decompose 
the phase so defined into a dynamical and a geometrical part was made in ref. \cite{jsrb}. 
The geometric phase for a noncyclic evolution was defined in this 
work as being equal to the integral of a two-form over the surface enclosed  by  the closed 
curve obtained by closing the 
open curve by ``any geodesic arc" connecting the final state to the initial state. 
For a two-state system this is equal to half the solid angle subtended by the closed 
surface at the center of the sphere.   There 
are two problems with this definition. Since the definition allows choice of any geodesic arc, 
if the open curve were closed by the longer 
geodesic arc connecting the final to the initial state, this definition gives a geometric 
phase which differs from the correct phase by $\pi$. Secondly, for systems more 
complicated than the two-state system, the projective Hilbert space is difficult to 
visualize hence this definition is  difficult to use. 

Following these developments, the present author, analyzing a gedanken 
polarization experiment \cite{jumps}, proposed an algebraic definition of the noncyclic geometric phase. 
 It was proposed in \cite{jumps} that the difference of the 
total phase as given by the Pancharatnam's criterion 
and the dynamical phase as defined by Aharonov and Anandan in \cite{aa} be defined as the 
geometric phase. It was stated that for a two-state system the geometric phase so defined 
equals half the solid angle of the closed curve obtained by closing the open curve with 
the shorter geodesic arc connecting the final state with the initial state.  In this 
work, the reference  beam was assumed to be in the same state as the initial state 
incident on the interferometer \cite{footnote2}. It was shown that the variation of the phase of the 
evolved beam as a function of parameters of the hamiltonian has several counter-intuitive features. 
In some regions of the parameter space the phase was found to be insensitive to variation in 
the parameters while in other regions the phase was found to vary sharply and discontinuously in the 
vicinity of points of singularity in the parameter space exhibiting $\pm \pi$ phase jumps. The shorter 
geodesic rule was shown to play an important role in understanding the counter-intuitive features.  
In a series of interference 
experiments involving polarization states of light \cite{dirac1,dirac2,4pism,iwbs} these 
features and the existence of phase singularities were explicitly demonstrated. In \cite{dirac1,dirac2} 
it was demonstrated that the total integrated phase shift measured in going around a circuit encircling 
phase singularities in 
the parameter space is equal to $2n\pi$ where $n$ is an integer and that it is zero for a circuit 
not enclosing a singularity. A review of these experiments and 
related work can be found in \cite{rbreview}. 

In view of recent interest in optics in noncyclic polarization changes and the Pancharatnam phase \cite{vandijk} we 
revisit the work first reported in  \cite{jumps} and present a generalization of the considerations in  \cite{jumps} to a scenario 
where both beams in the interferometer undergo unitary transformations and present a simple proof for the shorter 
geodesic rule in the more general context. This generalization provides a new degree of freedom 
to tune the location
of the phase singularities in potential applications. We then propose a formal definition of the nonmodular 
total topological phase, identify its dynamical and geometric components and note that in a cyclic variation of 
the parameters of the hamiltonian, the dynamical phase change integrates to zero so that the 
nonmodular total phase change over a cycle is equal to the geometric phase change, each being 
equal to $2n\pi$ with $n$ being an integer. 
Using a specific example of an SU(2) transformation in one arm of the
interferometer, the parameters of which are varied, while those in the other beam remain fixed,
we demonstrate by numerical sumulation  
discrete transitions between regions where $n$ has different values so that the global slope 
of the topological phase curve can change abruptly for small changes in a parameter. 
We then present an effective hamiltonian picture for the interpretation of phase changes 
in an interferometer with $SU(N)$ elements.

\section{The nonmodular topological phase}

Consider a version of the standard two-slit interference experiment shown in Fig. 1 in which 
a coherent plane wave of particles  with $N$ internal quantum states, propagating in the $z$-direction is 
incident on the two slits A and B. Let $|E>$ be the incident wavefunction, represented 
by a column vector with $N$ complex numbers. Let a box be placed in front of each of 
the slits such that passage through the box results in a unitary 
transformation $W_A$ or $W_B$ being applied to the wavefunction $|E>$,  changing 
its state to $|A>= W_A|E>$ or $|B>= W_B|E>$. $W_A$ and $W_B$ are functions 
of parameters ${\vec{\beta}}_A$ and ${\vec{\beta}}_B$ which can be varied during the 
experiment. Let a screen be placed 
some distance away where the two waves are superposed and the total intensity $I$
recorded at some point P on the screen. The states   at P are given by,
$|A> e^{i{\chi}_A}$ and $|B> e^{i{\chi}_B}$, where ${\chi}_A$ and ${\chi}_B$ 
are isotropic (state-independent) phases  acquired by the 
wavefunctions $|A>$ and $|B>$ during propagation to the point P.  Let $W_A= e^{i{\eta}_A}U_A$ 
and  $W_B= e^{i{\eta}_B}U_B$ where ${\eta}_{A,B}$ are isotropic phase factors associated with 
$W_{A,B}$ and $U_{A,B}$ are SU(2) transformations. The intensity $I$ at P 
is given by,

\begin{equation}
I~ =~ <A\mid A>  + <B\mid B>  ~+~  2 Re[<\tilde {B} \mid \tilde{ A}> e^{i({\eta}_A-{\eta}_B)} e^{i({\chi}_A-{\chi}_B)}], \label{eq:panch1}
\end{equation}

\noindent where $\mid \tilde{ A}> e^{i{\eta}_A}=\mid A>$ and $\mid \tilde{ B}> e^{i{\eta}_B}=\mid B>$.
The fringes on the screen are obtained due to variation of the 
phase $({\chi}_A-{\chi}_B)$ along the screen.
 The quantity $<\tilde {B} \mid \tilde{ A}> e^{i({\eta}_A-{\eta}_B)} e^{i({\chi}_A-{\chi}_B)}$ is the complex 
fringe visibility whose phase is the phase difference between the states 
$|A>$ and $|B>$  and contains a shift 
in the fringe pattern from its position in the absence of the unitary transformations $W_A$ and $W_B$, the latter being  given by
arg  $<E \mid E> e^{i({\chi}_A-{\chi}_B)}={\chi}_A-{\chi}_B$.
 This shift consists of two parts: (i) the phase of an isotropic, i.e. polarization-independent phase factor $ e^{i({\eta}_A-{\eta}_B)}$
and (ii) a polarization-dependent phase shift $\psi={\rm arg}<\tilde {B} \mid \tilde{ A}>$. We shall 
call (ii) the total Pancharatnam phase associated with the evolution.
 The 
points in the parameter space at which the 
magnitude of the fringe visibility $\mid<\tilde {B} \mid \tilde{ A}>\mid$ vanishes are the 
singular points where the contrast in the fringe pattern is zero and its phase is undefined.
The singular points could be isolated points or lines or more complicated surfaces in the 
parameter space. The phase shift $\psi$ is determined modulo $2\pi$. 
However if small 
changes in parameters ${\beta}_A$ and ${\beta}_B$ result in a  change $d\psi$ in the 
phase $\psi$, the integrated phase shift $\int {d\psi}$ is nonmodular. This is the 
quantity measured in the optical polarization implementations of the above experiment 
reported in (\cite{dirac1}-\cite{iwbs},\cite{nonintphase}).

The physical implementation of the unitary transformations $W_A$ and $W_B$ will depend 
upon the nature of the particle used in the interference experiment. For polarization 
states of light waves, these would typically be a birefringence, possibly variable, 
distributed along the length of the box represented by the variable  $s$. In case of 
spin-1/2 states of neutrons or atoms these could be  a magnetic field 
distribution, again possibly variable, along the length of the box. An infinitesimal 
unitary transformation in going from $s$ to $s+ds$,  is given by 

\begin{equation}
W_{A,B}(s,s+ds)=1-\frac{i}{\hbar} H_{A,B}(s)ds, \label{eq:hamiltonian}
\end{equation}

\noindent where $H_{A,B}$ (i.e. $H_A$ or $H_B$) is a hamiltonian function determined by the properties of the medium 
in the box represented by the parameters ${\beta}_{A,B}$. 

\noindent The finite unitary transformation $W_{A,B}$ can be formally written as,
 
\begin{equation}
W_{A,B}=e^{-\frac{i}{\hbar}\int H_{A,B}(s)ds}. \label{eq:unitary}
\end{equation}

\noindent

As the particle propagates through the box $A$ or $B$, the internal state of 
the particle will traverse some path in its state space, shown as $EA$ or $ESB$ in Fig.(2). Following Aharonov and Anandan \cite{aa}, we can define 
dynamical phases ${\psi_d}^A$ and ${\psi_d}^B$ for these evolutions as, 

\begin{equation}
{\psi_d}^A=-\frac{i}{\hbar} \int <A (s)\mid H_A (s)\mid A (s)>ds~~{\rm  and}
\label{eq:totdynphaseA}
\end{equation}
\begin{equation}
{\psi_d}^B=-\frac{i}{\hbar} \int <B (s)\mid H_B (s)\mid B (s)>ds.
\label{eq:totdynphaseB}
\end{equation}

\noindent We now define the geometric phase for the evolution as :
 
\begin{equation}
\psi_g=\psi - ({\psi_d}^A-{\psi_d}^B) , \label{eq:totgeomphase}
\end{equation}

\noindent where $\psi = <\tilde B\mid\tilde A>$ is the total Pancharatnam phase for the evolution, defined modulo $2\pi$.
For the special case when there is no evolution in  arm B, the above definition reduces to 
the definition given in ref.\cite{jumps}. For a two-state system the geometric phase $\psi_g$
so defined is equal to half the solid angle of the area EABSE in Fig (2) where AB is the 
shorter geodesic arc joining the points A and B.  This is the first main result in this paper. 

We now give a simple proof of the above shorter geodesic rule in the context of a two-state system. 
The proof consists of three steps:

1. Consider the two interference experiments shown in Figs. 1(a) and 1(b). In 1(a) the 
two interfering beams are in different states A and B. In 1(b) a polarizer $P_B$ that 
projects any state on state B is placed 
after the unitary transformation $W_A$ so that the two interfering beams are in the same 
state B. It is easy to show that the phase of the visibility in the two experiments is 
always the same. It is also well known that the action of the polarizer is to bring 
the state A to the state B along the shorter geodesic arc connecting A to B. 

2. Now remove the unitary transformation $W_B$ in front of slit B and place a unitary 
transformation ${W_B}^{\dagger}$ in front of the polarizer $P_B$ of experiment 1(b).
This is shown in Fig.1(c). It is again easy to see that the phase of the visibility 
in expts. 1(b) and 1(c) is always the same. 

3. Since the expt. of Fig.1(c) corresponds to a cyclic evolution of the state in arm A, 
one can use the decomposition of Aharonov and Anandan \cite{aa} to the closed path EABSE 
for which the dynamical phase is $({\psi_d}^A-{\psi_d}^B)$. It follows that $\psi_g$ given by 
 Eq.( \ref{eq:totgeomphase}) is equal to half the solid angle of the area enclosed 
by the curve EABSE.

It is conceptually simpler if the evolution in the interferometer were looked upon as a two-step 
process,  first,  in the interferometer free from SU(2) evolution,  the transformation $W_B$ 
is introduced in arm B and second, the transformation $W_A$ 
is introduced in arm A. The first stage is represented by the circuit ESBGE in Fig.2(a) and the 
geometric phase acquired by beam B equals half the solid angle of the area ESBGE. 
The geometric phase acquired by beam A in the second stage of the evolution must therefore 
be equal to half the solid angle of the area EABGE if the total geometric phase difference 
for the total evolution has to add up the value stated above. We therefore have the 
following very useful general result: If the reference beam in the interferometer is in state B
(doesn't matter how it gets there) and in the measurement beam the state evolves from E to A along the curve EA then the 
geometric phase acquired by beam A is equal to half the solid angle of the closed curve 
EABGE where AB is the shorter geodesic arc between A and B and BGE is the shorter 
geodesic arc between B and E. This is our second  main result in this paper. 

The relation between the geometric phase so defined and the more familiar quantities formulated 
by Berry \cite{berry} and by Aharonov and Anandan  \cite{aa} can be illustrated by the 
following example. Let a slab of a cholesteric liquid crystal with its helical axis 
along the direction of propagation be placed in the path of the beam. Let each layer in 
the crystal, in addition to linear birefringence, have some optical activity. The 
propagation of polarized light through such a medium is equivalent to the evolution 
the spin state of a spin-1/2 particle under the action of a rotating magnetic 
field, the field making a constant angle $\theta$ with the axis of rotation of 
the field (ref. \cite{rbreview}, p.48). It is well known that under such a hamiltonian, an 
arbitrary state does not undergo cyclic evolution but a special pair of orthogonal 
states do \cite{wang}. The Aharonov- Anandan phase relates 
only to the special pair of cyclic states. Berry's phase is a further special 
case when the rotation of the state on the state sphere due to birefringence of 
a single layer is large compared to the rotation of the birefringence axis 
from one layer to the next (adiabatic limit). 
For all states other than the cyclic states one needs the definition given 
by Eqn. (\ref{eq:totgeomphase}).

We now introduce the nonmodular phase. For small changes  in the parameters ${\vec{\beta_A}}$ ,
the phase changes $d \psi$, $d {\psi_d}^{(A,B)}$ and $d\psi_g$ are related by, 

\begin{equation}
d \psi= d {\psi_d}^A -  d {\psi_d}^B  +d\psi_g.\label{eq:totphase}
\end{equation}

\noindent  For finite changes in the 
parameters ${\vec{\beta_A}}$ , we therefore have 

\begin{equation}
\int d \psi= \int d {\psi_d}^A -  \int d {\psi_d}^B  +\int d\psi_g.\label{eq:totnonmodphase}
\end{equation}

\noindent The structure of Eqs.(\ref{eq:totdynphaseA}) and (\ref{eq:totdynphaseB}) makes it fairly obvious that
for any cyclic change in the parameters ${\vec{\beta_A}}$,

\begin{equation}
 \int d {\psi_d}^A=\int d {\psi_d}^B=0. \label{eq:statement2}
\end{equation} 

\noindent Eq. (\ref{eq:statement2}) implies that if one of the parameters ${\vec{\beta}_{A,B}}$ 
is an angle variable, ${\psi_d}^{A,B}$ must be a periodic function of this variable. 
 Now since a cyclic change in the parameters ${\vec{\beta}_{A,B}}$ must leave the fringe pattern 
unchanged,  Eqs.  (\ref{eq:totnonmodphase}) and (\ref{eq:statement2}) lead to the result

\begin{equation}
 \int d {\psi}_g = \int d\psi = \pm 2n\pi, \label{eq:statement3}
\end{equation}

\noindent where $n$ is an integer. By a cyclic change we mean that at the end of the change 
the box $W_A$ is physically identical to what it was before the change. Eq. (\ref{eq:statement3}) is 
the third main result in this paper. An example of a cyclic change is rotation 
in space through $2n\pi$ of any of the objects used to make the unitary transformation $W$  
or if the object has $m$-fold symmetry about some axis, rotation through $2n\pi/m$ about that axis. 
In case of the polarization experiments referred above, the objects used 
to make the unitary transformations $W_A$ are quarterwave and halfwave retarders 
which have a two-fold symmetry about the beam axis and the parameters 
${\vec{\beta_A}}$ are angles of rotation of these optical elements about the beam axis. In this case therefore Eqs.
(\ref{eq:statement2}) and (\ref{eq:statement3}) are true for rotations through $n\pi$ of any one or more of these optical elements.
For example in the experiment of ref. \cite{nonintphase},  rotation of a halfwave retarder 
about the beam axis through an angle $n\pi$ yields a total phase change equal to $2n\pi$.
A closed cycle in the space of parameters ${\vec{\beta_A}}$, where 
the net rotation of each optical element is zero, is a particularly interesting case where the 
Eqs. (\ref{eq:statement2}) and (\ref{eq:statement3}) are true. For example in the experiments reported in \cite{dirac1,dirac2} 
a closed circuit enclosing several phase singularities in the parameter space yielded 
a total phase change equal to $2 \pi \Sigma_k n_k$ where $n_k$ is $+1$ or $-1$
depending on the sign of the $k$ th singularity. Eqns. (\ref{eq:statement2}) and (\ref{eq:statement3})
then imply that these phase changes are geometric. 

A geometric picture for the nonmodular geometric phase shift would be as follows. Consider a typical 
interference situation where the reference beam has been brought to the state B somehow and one is interested 
in the phase shift as the parameters in beam A are varied. B can be any state on the sphere although in Fig. 2 it 
has been chosen to be the right circularly polarized state to be consistent with the specific example 
discussed in the next section. The arc $EA$ is a small circle traced by the state of beam A 
during passage through some SU(2) element placed in the beam. As stated above the geometric phase 
is equal to half the solid angle subtended by the closed curve 
$EABGE$ at the centre of the sphere, where $AB$ is the shorter geodesic arc joining the points 
$A$ and $B$  and $BGE$ is the shorter geodesic arc joining $B$ and $E$.  The nonmodular geometric phase shift is the integrated 
change in the shaded area  in Fig. 2a as the arc EA undergoes changes due to 
changes in the parameters of $W_A$.  With the help of Fig. 2b which shows the 
evolution of the arc $EA$ to the arc $EA'$ due to  changes in the parameters in 
beam A the swept area $S$ can be calculated as: 

\begin{eqnarray}
 S = EA'BGE-EABGE = (EA'QE+EQBGE)-(EQBGE+QABQ) \nonumber \\ 
 =EA'QE-QABQ =(EA'AE-A'AQA')-QABQ=EA'AE-BA'AB  \nonumber \\
=EA'AE+A'BAA'=~{\rm Area~ swept ~by~ the~ moving ~arc~} EA + ~~~~~~~~~~~~~\nonumber \\
{\rm Area~ swept~ by~ the~ moving~ geodesic ~arc}~ AB.   ~~~~~~~~~~~~~ \label{eq:ampphase}
\end{eqnarray}

As explained in \cite{jumps}, the phase jumps are understood 
as a sudden change in the  area swept on the sphere by $AB$ due to a sudden switch of the 
shorter geodesic arc near a point of singularity where the two interfering states are orthogonal. 
In many simple examples, the truth 
of Eq. (\ref{eq:statement3}), hence of Eq. (\ref{eq:statement2}) can be verified in terms of this geometric picture.

\section{A specific example}

Consider an interference experiment with particles with two internal states, say 
 the two polarization states of a photon or the two spin states of a neutron. 
Let us choose an orthogonal set of circularly polarized  
states (or $|\pm z>$ states in case of spin-1/2 particles)  as the basis states and choose the relative phase between them 
to be such that the state $(1/\sqrt 2)[|+z>+|-z>]$ corresponds to a linear 
polarization along the $x$-axis (or to the spin-1/2 state $|x>$).  Let a state $|E>$=$(1/\sqrt 2)[|+z>+|-z>]$ be incident on 
the interferometer. Let the reference beam B be brought to the right circularly polarized state by means of 
a quarterwave plate or a circular polarizer so that 

\begin{eqnarray}
 |B>=(1/\sqrt 2)|+z>  \label{eq:refstate}
\end{eqnarray}

Let an SU(2) element corresponding to a rotation through an angle $\delta$ (also called retardation)
about an axis in the direction represented by the point $(\theta,\phi)$ on the 
state sphere be placed in arm A of the interferometer so that 

\begin{eqnarray}
W_A={\rm cos}(\delta/2)-i{\rm sin}(\delta/2)[\vec\sigma.\hat n], \label{eq:su2element}
{\rm where}~ \hat n=({\rm sin}\theta {\rm cos}\phi,{\rm sin}\theta {\rm sin}\phi,{\rm cos}\theta).
\end{eqnarray}

\noindent The  components $\sigma_x,\sigma_y$ and $\sigma_z$ of the vector $\vec\sigma$ are the three Pauli matrices.  The final state $|A>$ after passage of the beam through 
arm A is given by, 

\begin{eqnarray}
 |A>=W_A|E> \label{eq:finalstate}
\end{eqnarray}

\noindent A simple calculation shows that the complex visibility $V$ of the interferometer 
is given by,

\begin{eqnarray}
 V=<B|A>=[{\rm cos}(\delta/2)-{\rm sin}(\delta/2){\rm sin}\theta {\rm sin}\phi]
-i{\rm sin}(\delta/2)[{\rm cos}\theta+{\rm sin}\theta{\rm cos}\phi]. \label{eq:visibility}
\end{eqnarray}

\noindent If $v$ and $\psi$ represent the amplitude and phase of $V$, these quantities can be 
separately determined uniquely from the equations

\begin{eqnarray}
 v_r=v{\rm cos}\psi={\rm cos}(\delta/2)-{\rm sin}(\delta/2){\rm sin}\theta {\rm sin}\phi\nonumber \\ 
~{\rm and}~ v_i= v{\rm sin\psi}=-{\rm sin}(\delta/2)[{\rm cos}\theta+{\rm sin}\theta{\rm cos}\phi] \nonumber \\
{\rm so ~that}~~ {\rm tan}\psi=(v_i/v_r)\label{eq:ampphase}
\end{eqnarray}

\noindent Consider now an experiment in which $\delta$ and $\theta$ are held fixed and the SU(2) element 
is rotated about the beam axis from $0$ to an  angle $\phi/2$ so that its azimuth on the state sphere rotates by 
an angle $\phi$. The nonmodular total phase shift is given by 

\begin{eqnarray}
 \int d\psi =\int[(v_rdv_i-v_idv_r)/({v_i}^2+{v_r}^2)].  \label{eq:totphaseshift}
\end{eqnarray}

\noindent The initial phase shift ${\psi}_0$ is obtained by substituting $\phi=0$ in Eqs.(\ref{eq:ampphase}) 
and computing the argument using the inverse trigonometric function.

The dynamical phase acquired by the beam in passing through the SU(2) element is given by \cite{aa}

\begin{eqnarray}
 {\psi_d}=-(\delta/2){\rm cos}\alpha, \label{eq:dynphase1}
\end{eqnarray}

\noindent where $\alpha$ is the angular length of the geodesic arc connecting the points $(90^\circ,0^\circ)$  and 
$(\theta,\phi)$ on the sphere, given by,

\begin{eqnarray}
 {\rm cos}(\alpha/2)=|<90^\circ,0^\circ|\theta,\phi>| \label{eq:arclength}
\end{eqnarray}

\noindent Eqs. (\ref{eq:dynphase1}) and (\ref{eq:arclength}) give

\begin{eqnarray}
 {\psi_d}=-(\delta/2){\rm sin}\theta {\rm cos}\phi. \label{eq:dynphase2}
\end{eqnarray}
\noindent Note that $\psi_d$ is a periodic function of the angle variable $\phi$.

We have computed the nonmodular total phase shift $\psi$ which is the sum of ${\psi}_0$ and the 
quantity given by Eq.(\ref{eq:totphaseshift}), for $\theta=90^\circ$ and various values of $\delta$ as 
a function of $\phi$. The hamiltonian acting on the beam in this case is exactly the same as 
that in the second example discussed in the previous section.
The angle $\phi$ equals twice the angle of rotation of the SU(2) element about the beam axis.
This corresponds to the second example considered in the previous section.  
Figs. (3-6) show the results. The polar angle $\theta$ of the eigenstate of the SU(2) element is 
equal to $90^\circ$ i.e. it lies on the equator in all the cases.  The interesting thing 
to note is the behaviour of the phase curves in the vicinity of the points $(\delta=90^\circ,\phi=90^\circ)$, 
$(\delta=90^\circ,\phi=450^\circ)$ in Fig. 4 and the points 
$(\delta=270^\circ,\phi=270^\circ)$, $(\delta=270^\circ,\phi=630^\circ)$ 
in Fig. 6. These are the points where the two interfering beams are in orthogonal states. 
The phase jumps by $+\pi$ or $-\pi$ depending on whether $\delta$ is less than or greater than 
$90^\circ$ in case of Fig. 4 and on whether $\delta$ is greater than or less than 
$270^\circ$ in case of Fig. 6. Also note the change in the global slope of the phase curve 
for a small change in $\delta$ across the value $90^\circ$ in Fig. 4 and the value $270^\circ$ 
in Fig. 6. The curve for $\delta=180^\circ$ 
shown in Fig. 5 shows a linear variation of the phase of the beam as a function of rotation 
of the SU(2) element. This represents a pure frequency shift of the beam equal to twice the 
rotation frequency of the SU(2) element. A nonmodular phase curve with a nonzero global 
slope is an unambiguous signature of a topological phase and was first demonstrated in 
an interference experiment some time ago \cite{nonintphase}.

The dynamical phase as a function of $\phi$, as given by Eq. (\ref{eq:dynphase2}), is shown in Fig. 7 
for $\delta=180^\circ$. The curves for other values of $\delta$, being exactly similar, with amplitude 
$\delta/2$, are not shown separately. The global slope of this curve is always zero. 
The geometric part of the nonmodular
topological phase, as defined in \cite{jumps} is the difference of the total phase as shown in 
Figs.(3-6) and the corresponding dynamical phase curve. Since the dynamical phase is featureless 
all the interesting features of the topological phase can be traced to the geometric part of the 
phase. However since the total phase contains all the 
interesting features, the curves for the geometric phase are not displayed separately.

\section{The effective hamiltonian picture}

The time evolution of the state of the beam at the point of interference can be described in 
terms of an effective hamiltonian $H_{eff}(t)$ as follows.

Let us assume  that  $W_A=W(t)$ where the 
time variation of $W(t)$ comes from time variation of its parameters $\vec {\beta}$.
If the incident state  is $\mid E(t)=\mid E(0)>$,
the state $\mid A (t)>$ after the SU(2) transformations
is  given by,
\begin{equation}
\mid A (t)> ~= ~W(t) ~\mid E (t)> ~= ~W(t) ~\mid E (0)>\label{eq:wt}.
\end{equation}

so that,
\begin{eqnarray}
\mid A (0)> ~ = ~ W(0) ~\mid E (0)> ~ \label{eq:psiitof}.
\end{eqnarray}

\noindent Now we can define a time-evolution operator $U(t)$ governing the evolution of
$\mid A(t)>$ as,
\begin{equation}
\mid A (t)> ~= ~U(t) ~\mid A (0)> \label{eq:utdef}.
\end{equation}

\noindent From the above equations, it follows that,
\begin{equation}
U(t) ~= ~W(t) ~ W^{\dagger}~(0)   \label{eq:uw}.
\end{equation}

\noindent Eqns. (\ref{eq:utdef}) and (\ref{eq:uw}) give,
\begin{equation}
\mid A (t+\delta t)>~=~U(t+\delta t) ~ U^{\dagger} (t) ~\mid A (t)>
~=~ W(t+\delta t) ~ W^{\dagger} (t) ~\mid A (t)> \label{eq:smallevol}.
\end{equation}

An effective hamiltonian $H_{eff}$ for the evolution of
$\mid A (t)>$ can now be defined by the relation:

\begin{equation}
exp[-(i/\hbar) ~H_{eff}(t) ~\delta t]  ~=~W(t+\delta t)~ W^{\dagger} (t) \label{eq:effh}
\end{equation}

\noindent It follows that
 
\begin{equation}
1-(i/\hbar)H_{eff}(t) ~\delta t~=~W(t+\delta t)~ W^{\dagger} (t)
\end{equation}

\noindent This leads to,

\begin{equation}
H_{eff}(t) = i\hbar \dot{W}(t)~ W^{\dagger} (t) \label{eq:heff}
\end{equation}

\noindent Now if $W(t)$ is cyclic, i.e. if $W(t+T)=W(t)$, the hamiltonian generated 
by eqn.(\ref{eq:heff}) leads to a cyclic $U(t)$, i.e. $U(T)=U(0)=\bf{1}$ which in turn 
implies that all states reproduce after time $T$ along with their phases which must 
therefore be equal to $2n\pi$. The value of $n$ will also depend on the reference state 
used for interference.
This general recipe is valid for $N$-state systems 
where $W(t)$ is an SU($N$) matrix and can be used as a recipe to generate hamiltonians 
under which entire state spaces undergo cyclic evolution. 

We illustrate the effective hamiltonian picture with two  examples for $N$=2. 
As the first example, let a monoenergetic beam of spin-1/2 particles propagate through 
a box with a magnetic field along $\hat z$ whose magnitude varies 
linearly with time.  Let the magnitude of the field and the length L of the box be adjusted 
so that the particles precess through an angle $\delta=at$ during passage through the box.
$W(t)$ for this setup is given by, 

\begin{equation}
W(t)={\rm cos}(at /2){\bf 1}-i{\rm sin}(at /2){\sigma}_z
\end{equation}

\noindent Use of eqn.(\ref{eq:heff}) then leads to the effective hamiltonian matrix:

\begin{equation}
H_{eff} = (a\hbar/2){\sigma}_z
\end{equation}

At the point of interference therefore the evolution of the state is like precession
in a constant magnetic field. If a beam of spin-1/2 particles with the spin state making 
an angle $\theta$ with the z-axis is incident on the interferometer and the initial state 
is taken as the reference state, the measured phase variation as a function of time will 
be as shown in Fig. 3  with the curves for $\theta=89.99$ and $\theta=90.01$
showing $\pm \pi$ phase jumps at values of $t$ for which the  interfering states become 
orthogonal. Fig. 3 is essentially the same as Fig. 4 of ref.\cite{4pism}. We note that the sign of the $\pi$- phase jump is measurable 
and has indeed been measured in \cite{4pism,iwbs} where both $+\pi$ and $-\pi$ phase jumps 
are seen.  Some optical interference experiments by other groups \cite{schmitzer,li} in similar contexts have measured 
phase jumps of only one sign leaving behind the impression that the sign is insignificant. 
We emphasize that a full description of the singularity requires both $+\pi$ and $-\pi$ phase jumps.
A proposal to observe phase singularities in a neutron interferometer experiment was made 
in \cite{neutronifproposal}.

We may also mention in passing that 
instead of spin-1/2 particles, if a beam of 
spin-$N$ particles were prepared in an initial state $\mid N,N>$ with the axis of quantization being 
along $\hat n=({\rm sin}\theta,0,{\rm cos}\theta)$ the observed phase curves will be exactly $N$ times 
those shown in Fig.3 with phase jumps equal to $\pm N\pi$. This nonmodular 
aspect is directly related to the total spin quantum number $N$ and is clearly physically significant.

As the second example, let a monoenergetic beam of spin-1/2 particles propagate through 
a box in the z-direction and a rotating magnetic field $\vec{B}=(B{\rm cos}\omega t,B{\rm sin}\omega t,0)$
be applied in the $x,y$ plane. Let the magnitude of the field and the length L of the box be adjusted 
so that the particles precess through an angle $\delta$ during passage through the box. 
This example corresponds  to the example for which the numerical simulations have 
been considered in the previous section. 
$W(t)$ for this setup is given by, 

\begin{equation}
W(t)={\rm cos}(\delta /2){\bf 1}-i{\rm sin}(\delta/2)({\rm cos}\omega t {\sigma}_x+{\rm sin}\omega t {\sigma}_y)
\end{equation}

\noindent Use of eqn.(\ref{eq:heff}) then leads to the hamiltonian matrix:

\begin{eqnarray}
H_{eff} = -\hbar \omega {\rm cos}\eta (\hat{n}.\vec{\sigma}),~~~~ {\rm where} \label{eq:defheff}\\
\eta=(90^\circ - \delta/2) \label{eq:defeta}\\
 {\rm and}~\hat{n}=({\rm sin}\eta {\rm cos}(\omega t+\pi /2),{\rm sin}\eta {\rm sin}(\omega t+\pi /2),{\rm cos}\eta ) \label{eq:cycliccond1}
\end{eqnarray}

At the point of interference therefore beam A sees an effective rotating magnetic field whose 
parameters are related to those of the rotating field placed in the path of the beam by Eqs.(\ref{eq:defheff}-\ref{eq:cycliccond1}).
For the special case  of $\delta=\pi$, i.e. for a rotating halfwave plate or a ``$\pi$-flipper" placed in 
the path of beam $A$, the equations give a constant effective field along $\hat z$ which is intuitively obvious.
Eqs.(\ref{eq:defheff}-\ref{eq:cycliccond1}) imply that for  a rotating magnetic field whose magnitude corresponds to  a Larmor 
precession frequency $\omega_L$,  which makes a constant angle  $\eta$  with the z-axis and which 
rotates about the z-axis with frequency $\omega$, the entire state space undergoes cyclic evolution 
when the following condition is satisfied by the parameters $\omega_L, \omega$ and $\eta$:

\begin{equation}
\omega_L =2\omega {\rm cos}\eta \label{eq:cycliccond2}
\end{equation}
\noindent As an illustration of the significance of this condition, in the example of the cholesteric 
liquid crystal discussed earlier, when the parameters 
of the sample satisfy the condition corresponding to Eq.(\ref{eq:cycliccond2}), the sample behaves, 
with respect to polarization, as a  piece of plain glass.

We note that while the total phase shift obtained in both the pictures is obviously the 
same the decomposition into a dynamical and a geometric phase is different. In the effective 
hamiltonian picture, the total nonmodular
phase shift is given by $\int d\psi(t)$ where 

\begin{equation}
\psi(t) = {\rm arg} <B\mid U(t)\mid A(0)>, 
\end{equation}

\noindent the dynamical phase is given by Eq.(\ref{eq:totdynphaseA}) with $H_A$ replaced by $H_{eff}$ and  
the nonmodular geometric phase is given by the area 
swept by the moving geodesic arc AB alone.

\section{Discussion}

In this paper we have explored a new dimension of the noncyclic geometric phase acquired by an 
evolving state namely its dependence on the reference state used for interference. This is like 
studying the phase of matrix elements of the time evolution operator $U(t)$ other than the diagonal 
matrix element studied earlier.

We have presented two different ways of understanding nonmodular phase changes in a generic interference 
setup shown in Fig.1, 
where the  unitary transformation $W_A$ is a  function of some variables $\vec\beta$ which can 
be varied in time. In the first picture one considers evolution through the sequence of 
elements whose product is $W_A$ at time $t$, computes the total Pancharatnam phase shift, 
does the same thing at a slightly later time  $t+\delta t$ and takes the difference to obtain the observed phase 
shift for the infinitesimal evolution. This is equal to the sum of (i) change in the dynamical phase shift and (ii) the change in the 
geometric phase shift which is given by the appropriate areas swept on the sphere.  Integration 
of such phase shifts give the total phase shift for the finite evolution.
In the second picture one constructs an effective time varying hamiltonian as 
seen by the beam at the point of interference and considers the phase shifts acquired 
by the final state for evolution under the effective hamiltonian. While the total phase 
shift obtained in the both the pictures is the same, the decomposition into the 
dynamical and the geometric parts is different .

While the second picture 
is  closer to the usual theoretical treatements, the first picture gives useful insights 
in cases where the evolution of the parameters in the main beam is cyclic. In such cases 
the dynamical 
phase shift $\int d{\psi}_d$ for a full period of $\vec\beta$ is equal to zero. In other 
words the nonmodular dynamical phase shift is a periodic function of $\vec \beta$ and its 
global slope is equal to zero. The  statements imply that when a beam of particles 
with $N$ internal states passes through an object which performs a sequence of $U(N)$ 
transformations on the state, the total nonmodular dynamical phase acquired by the state is 
defined unambigously for every incident state and is independent of a reference state. 
It is thus an intrinsic property of the system.
The nonmodular total phase shift and therefore the 
geometric part of the phase shift however need not integrate to zero over a period of $\vec\beta$ 
and can equal $2n\pi$ where $n$ is an integer. These quantities therefore need not be 
periodic functions of $\vec \beta$.  
Therefore, unlike the dynamical phase, the geometric phase acquired by the state is defined only modulo $2\pi$ 
and can change by $2N\pi$ under cyclic evolutions of the system, the value of $N$ 
depending not only on the cycle of parameter changes but also on the reference state 
with respect to which the phase changes are measured. For two-state systems such changes  are 
determined entirely by the appropriate areas swept on the sphere.

We have presented an example demonstrating that 
the total phase shift over a cycle of a cyclic parameter is absolutely robust except at singular points 
in  the parameter space where the nonmodular 
phase shift can make a sudden transition from one value of $n$ to another for a small change in parameter, i.e. can undergo
a discrete jump in its global slope. The phase variations can be extremely sharp in the 
vicinity of these points.

Both the above mentioned properties namely robustness of the global slope of the topological 
phase shift and the possiblity of discrete transitions from one value of $n$ to another can 
form the basis of applications perhaps in the area of quantum information. An example of the 
first kind is the 
achromatic retarders developed by Pancharatnam \cite{panch2} for polarized light. Some 
examples of the second kind  have been described in \cite{qhqjumps} where 
it is shown that (i) an array of radio antennas phased using topological phase shifters 
can be made to look in two different directions at two different wavelengths at the same 
time and (ii) one can make a geometric phase lens which can switch from being convex to 
being concave with a change of wavelength of light passing through it.

We also wish to note that while the focus in this paper is on unitary transformations, 
we expect the results to have suitable generalizations to nonunitary transformations.
Finally we point out an interesting extension of the Pancharatnam phase criterion  to 
partially polarized waves by Sj\"{o}qvist et. al  \cite{sjoqvist}. It was shown 
in \cite{comment} that phase singularities form an important part of 
the description in this case too.\\

\section{Note added}
The author's attention has been drawn by one of the referees to a paper by Berry \cite{berry2}  that   describes the theory of $2n\pi$ phase shifts arising from change in the number  of dislocation lines threading an interferometer. This paper makes the very interesting observation that even the isotropic, polarization-independent phase shifts arising from a change in the optical path difference between  two beams of an interferometer are topological in one sense.  The considerations in \cite{berry2} however differ from those in the present paper and in our earlier work in the following respects. While in \cite{berry2} the interferometer configurations corresponding to the different values of $n$ look physically different in that they enclose different number of dislocation lines, in the present paper and in our earlier work the $2n\pi$ phase shifts arising from cyclic variation of parameters in the interferometer beams leave the interferometer physically unchanged. Secondly, while the singularities in Berry's work reside in  physical space, those in our work reside in the space of parameters of the unitary transformations that act on the internal degrees of freedom of the interfering particle, for example polarization in case of the photon.

The relationship of our work with spatial singularities of the electromagnetic field can be further illustrated with the help of the following illustration. Let $\hat z$ be the propagation direction. At $z=-\epsilon$, let a screen be placed normal to the beam which is a polarizer that passes the state $|E>$. At $z=o$, let an "SU(2) screen" be placed which is such that at each point in the $x,y$ plane an element is placed which performs an SU(2) transformation on the fields that pass through that  point. At $z=\epsilon$, let a third screen be placed which is a polarizer that passes the state $|B>$. Now let the amplitude and phase of the fields at each point in a plane normal to the beam at $z=2\epsilon$ be examined. There will be points or lines at which the amplitude of the field will be zero. These are the singular points or lines. Now let a phase detector be taken in a closed circuit around an isolated singular point. If the phase detector is such that it integrates phase shifts as it goes along, the total phase shift recorded in one such circuit will be equal to $2n\pi$ where $n$ is the strength of the singularity. Such an experiment is not very convenient to perform. What can easily be done however is to replace $x and y$ with two variable SU(2) parameters in one of the beams of an interferometer such that variation of these parameters results in the same sequence of SU(2) transformations on the polarization state as in the spatial example above. This is what is done in our experiments and the two are equivalent. While the spatial analogy presented above works only for two parameters, in our experiments one can in principle have more than two variable parameters. 

\section{Acknowledgements}
A good part of the work reported in this paper was done during the period the author was 
employed with the Raman Research Institute, Bangalore. Library support from the institute after the author's retirement is gratefully acknowledged.

\newpage


\begin{figure*}
\includegraphics[height=150mm]{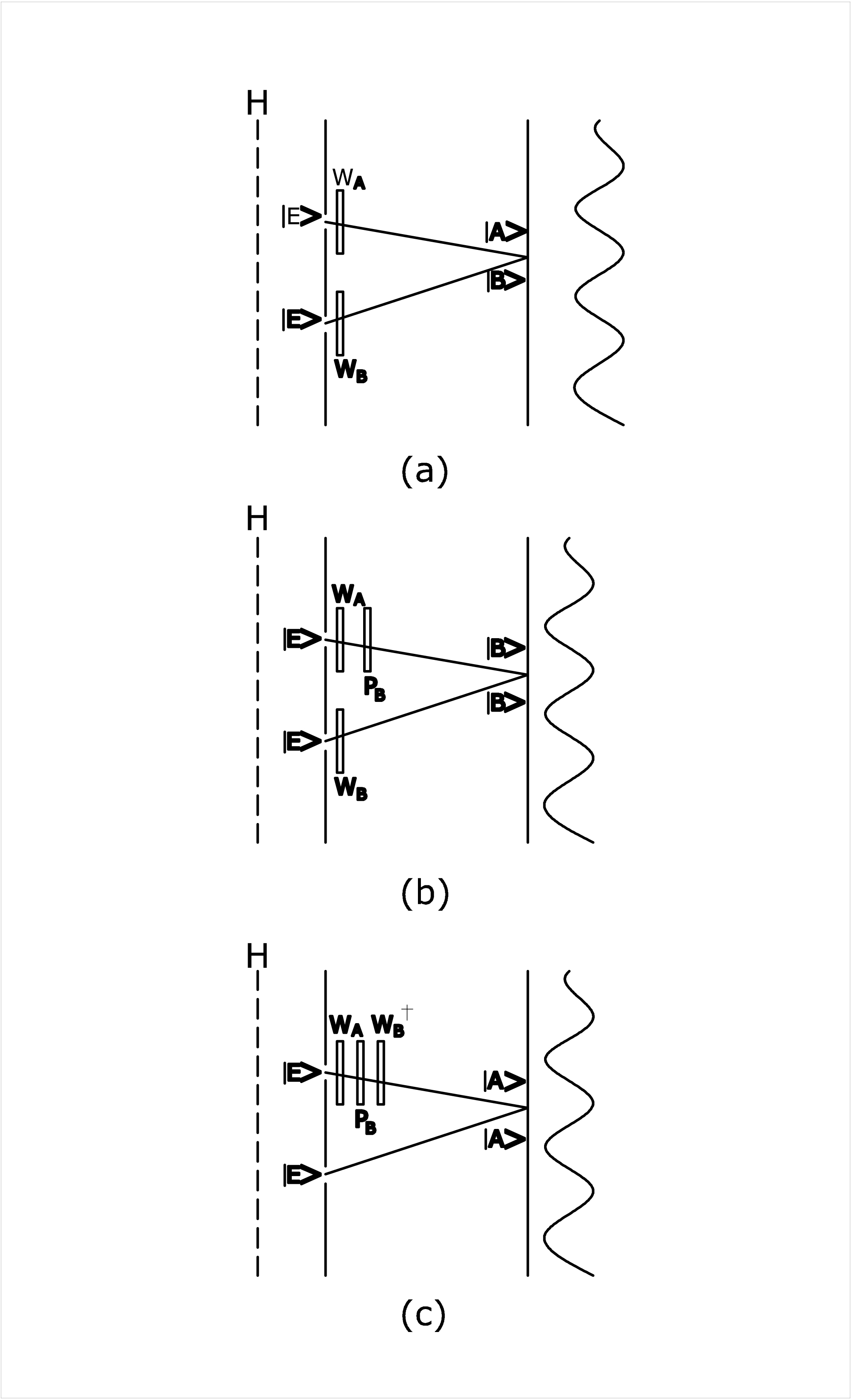}
\label{Fig.1}
\caption{(a) Interference pattern formed on a screen due to interference of two parts of a 
wavefront H in polarization states $\mid A>$ and $\mid B>$ produced by the action of unitary transformations 
$W_A$ and $W_B$ on the initial state $\mid E>$. (b) The phase of the visibility in expt. (a) remains unchanged 
if a polarizer $P_B$ that  brings any state to the state $\mid B>$ is placed in front of $W_A$ so that both 
interfering states are $\mid B>$. (c) If $W_B$ is removed and a unitary transformation ${W_B}^\dagger$ is 
placed in front of the polarizer $P_B$ in arm A, the phase of the visibility still remains unaltered. }
\end{figure*}

\begin{figure*}
\includegraphics[height=115mm]{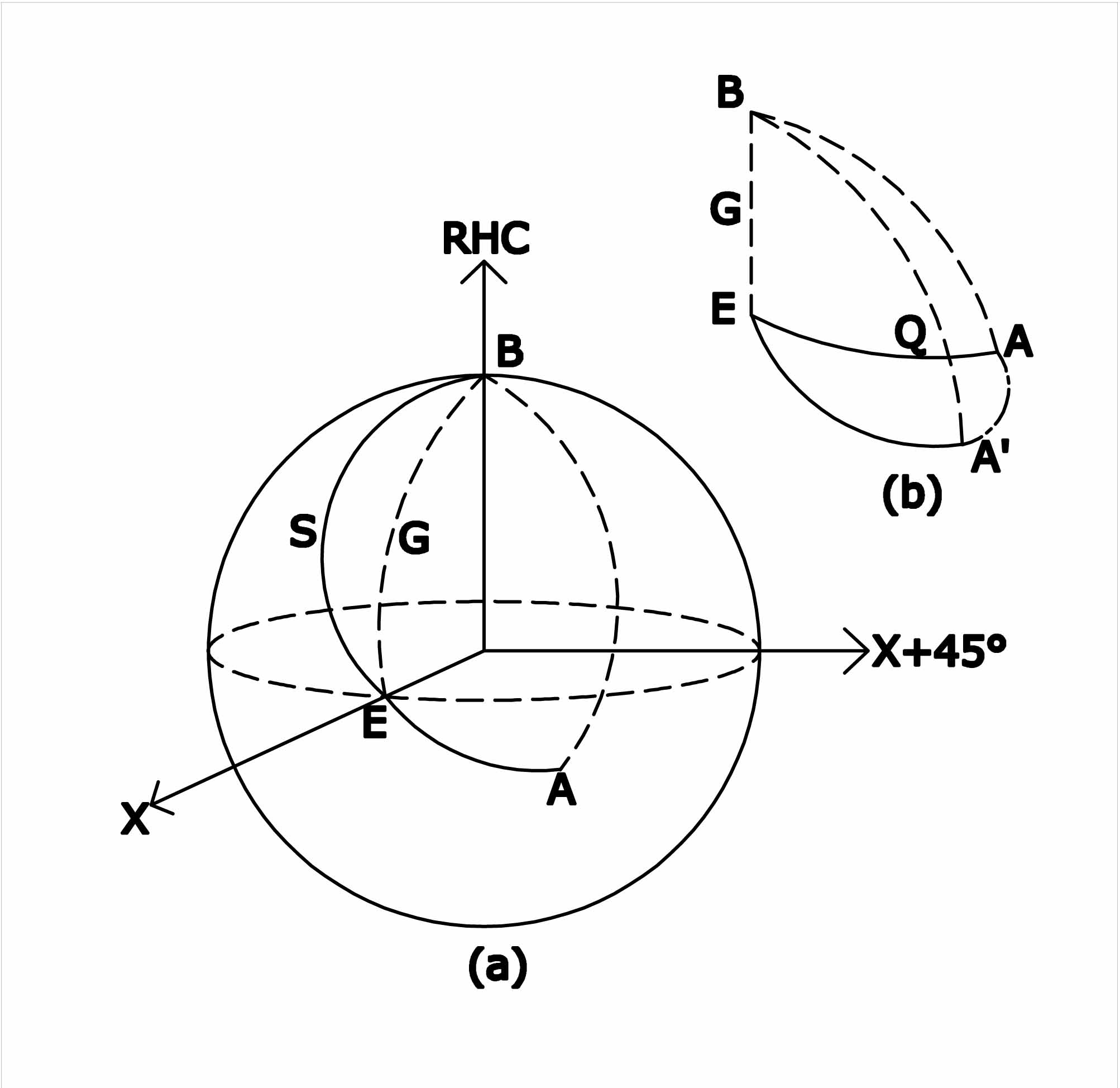}
\label{Fig.2}
\caption{(a) A geometric representation on the Poincar\'{e} sphere of transformations of 
the state of the 
two-state system of  corresponding to the experiment shown 
in Fig. 1(a). $EA$ is a small circle arc corresponding to the track of the polarization state 
due to the unitary transformation $W_A$ in arm A and  $ESB$ is a small circle arc corresponding to the track of the  state |B>
due to  $W_B$ in arm B. AB is the shorter geodesic arc connecting points A and B and 
BGE is the shorter geodesic arc connecting points B and E.  The geometric phase 
is equal to half the solid angle subtended by  the closed curve 
$EABSE$ at the centre of the sphere. This can be broken into two parts, the area ESBGE due to the transformation $W_B$ 
and the area EABGE  due to the transformation $W_A$.  (b) The change in the geometric phase as the 
small circle arc EA moves to EA' due to change in parameters of the unitary 
transformation in beam A is given by the difference in the areas EA'BGE and EABGE 
which can be shown to be equal to the area swept by the moving arc EA plus 
that swept by the moving geodesic arc AB.}
\end{figure*}

\begin{figure*}
\includegraphics[height=145mm]{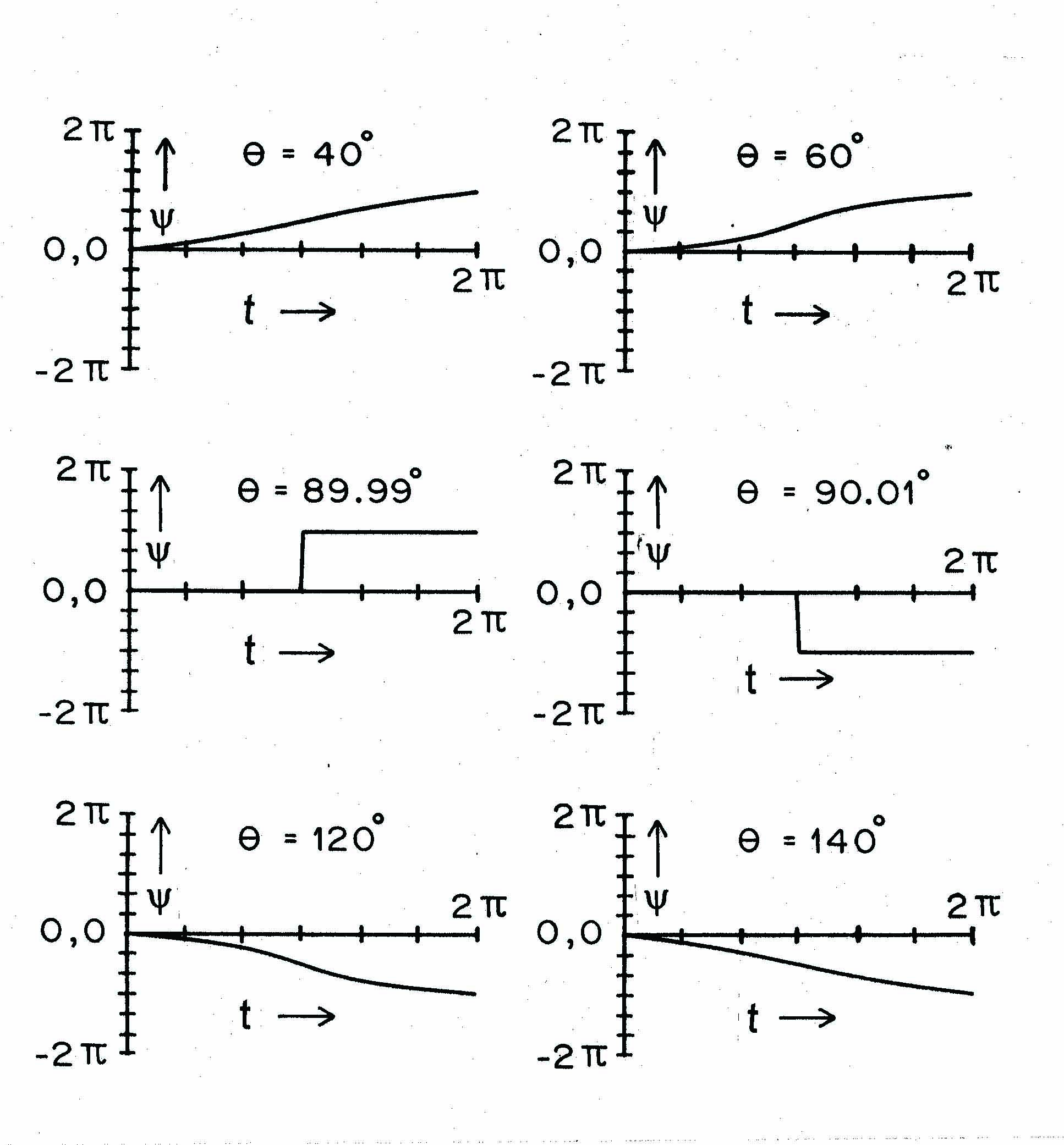}
\label{Fig.3}
\caption{Phase shift $\psi$ as a function of time t when the beam in arm $A$ passes through  
a magnetic field 
along $\hat z$ varying linearly with time; the reference state for interference being the same as 
the incident state. The variable $\theta$ represents the polar angle of the 
incident state on the state sphere and time is expressed as the angle of 
precession of the spin state. Note the singular behaviour of the phase shift 
in the vicinity of $\theta=90^\circ$ and precession angle $\pi$. For the 
successive precession periods, the curves repeat themselves with the value at the 
end of the period as the new zero for the phase.}
\end{figure*}

\begin{figure*}
\includegraphics[height=75mm]{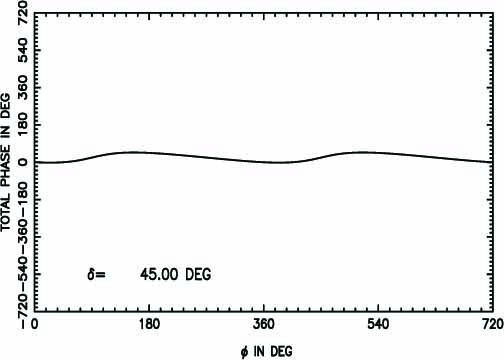}
\label{Fig.4}
\caption{The nonmodular total topological phase as a function of rotation of 
the SU(2) element on the Poincar\'{e} sphere when the retardation of the element is $45^\circ$. }
\end{figure*}

\begin{figure*}
\includegraphics[height=75mm]{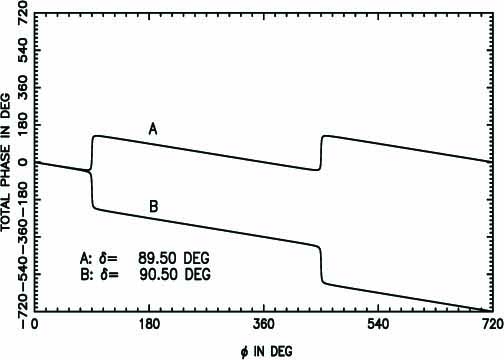}
\label{Fig.5}
\caption{The upper curve shows the nonmodular total topological phase as a function of rotation of 
the SU(2) element on the Poincar\'{e} sphere when the retardation $\delta$ of the element is $89.5^\circ$ and the lower curve 
shows the same quantity when the retardation  is $90.5^\circ$. Note the discrete $\pm \pi$ jump in the phase 
when the azimuth $\phi$ is close to $90^\circ$ or $450^\circ$ resulting in a discrete change in the global slope of 
the phase curve. The points ($\delta=90^\circ,\phi=90^\circ, 450^\circ$) are 
phase-singular points.}
\end{figure*}

\begin{figure*}
\includegraphics[height=75mm]{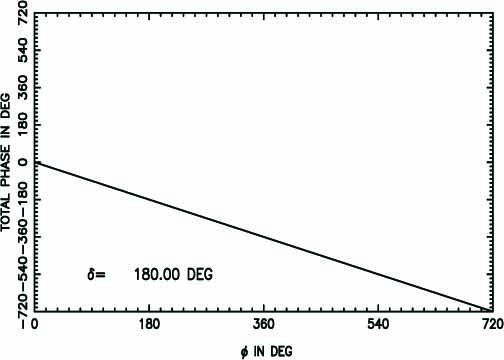}
\label{Fig.6}
\caption{The nonmodular total topological phase as a function of rotation of 
the SU(2) element on the Poincar\'{e} sphere when the retardation of the element is $180^\circ$.}
\end{figure*}

\begin{figure*}
\includegraphics[height=75mm]{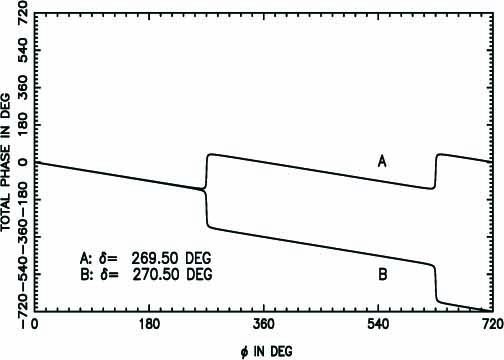}
\label{Fig.7}
\caption{The lower curve shows the nonmodular total topological phase as a function of rotation of 
the SU(2) element on the Poincar\'{e} sphere when the retardation $\delta$ of the element is $269.5^\circ$ and the upper curve 
shows the same quantity when the retardation is $270.5^\circ$. Note the discrete $\pm \pi$ jump in the phase 
when the azimuth $\phi$ is close to $270^\circ$ or $630^\circ$ resulting in a discrete change in the global slope of 
the phase curve. The points ($\delta=270^\circ,\phi=270^\circ, 630^\circ$) are 
phase-singular points.}
\end{figure*}

\begin{figure*}
\includegraphics[height=75mm]{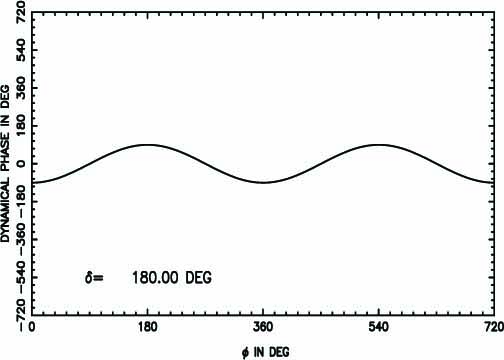}
\label{Fig.8}
\caption{The nonmodular dynamical phase for retardation $180^\circ$ as a function of rotation of 
the SU(2) element on the Poincar\'{e} sphere which is equal to half the angle of rotation of 
the element in real space.}
\end{figure*}

\begin{thebibliography}{4pism}
\bibitem{footnote2}
 The  algebraic definition for the noncyclic geometric phase introduced in \cite{jumps}  can be found incorporated in a later, more 
formal treatment of the subject by Mukunda and Simon \cite{muksim}. We also refer the theoretically inclined reader to a paper by Pati \cite{pati} that deals with the geometric aspects of noncyclic quantum evolution.
\end{thebibliography}
\end{document}